\documentclass[twocolumn,,12pt]{aastex631}

\usepackage{lineno}
\usepackage{epsfig}
\usepackage{natbib}
\usepackage{epstopdf}
\usepackage{mathrsfs}
\usepackage{bm}
\usepackage{slashed}
\usepackage{verbatim}
\usepackage{graphicx}
\usepackage{amssymb}
\usepackage{psfrag}
\usepackage{array}
\usepackage{lipsum}
\usepackage{float}
\usepackage{babel}

\begin{document}
\title{Deep Learning the Forecast of Galactic Cosmic-Ray Spectra}

\author{Yi-Lun Du}
\address{Shandong Institute of Advanced Technology, Jinan 250100, China}
\email{yilun.du@iat.cn}
\author{Xiaojian Song}
\address{Shandong Institute of Advanced Technology, Jinan 250100, China}
\author{Xi Luo}
\address{Shandong Institute of Advanced Technology, Jinan 250100, China}

\vspace{10pt}

\begin{abstract}
We introduce a novel deep learning framework based on Long Short-Term Memory (LSTM) networks to predict galactic cosmic-ray spectra on a one-day-ahead basis by leveraging historical solar activity data, overcoming limitations inherent in traditional transport models. By flexibly incorporating multiple solar parameters, such as the heliospheric magnetic field, solar wind speed, and sunspot numbers, our model achieves accurate short-term and long-term predictions of cosmic-ray flux. The addition of historical cosmic-ray flux data significantly enhances prediction accuracy, allowing the model to capture complex dependencies between past and future flux variations. Additionally, the model reliably predicts full cosmic-ray spectra for different particle species, enhancing its utility for comprehensive space weather forecasting. Our approach offers a scalable, data-driven alternative to traditional physics-based methods, ensuring robust daily and long-term forecasts. This work opens avenues for advanced models that can integrate broader observational data, with significant implications for space weather monitoring and mission planning.
\end{abstract}

%
%
%
%
%

\section{Introduction} 

Cosmic rays, high-energy particles originating from outside our solar system, pose considerable risks to both human health and electronic systems~\cite{cucinotta2014space,simonsen2020nasa,de2020crater,chen2023astronaut}. Primary cosmic rays and the secondary particles they produce upon interacting with the Earth's atmosphere contribute to elevated radiation exposure for astronauts, aviation crews, and high-altitude populations. Moreover, cosmic rays cause ionization, which can lead to single-event upsets in electronic devices, affecting the reliability of satellite operations and ground-based electronics. Given these concerns, accurate forecasting of cosmic ray intensities is crucial for various sectors, including aerospace, telecommunications, and defense.

As galactic cosmic rays (GCRs) propagate through the heliosphere, they interact with the solar wind and the frozen-in interplanetary magnetic field. Consequently, the flux of GCRs observed near Earth is modulated by solar activity~\cite{parker1965passage,quenby1984theory,simpson1998brief,luo2011cosmic,kota2013theory,luo2013galactic,potgieter2013solar,potgieter2014very,o2015badhwar,vos2015new,vos2016global,luo2017numerical,potgieter2017global,shen2019modulation,luo2020numerical,aguilar2021periodicities,song2021numerical}. Solar activity, which includes various events, such as solar flares, coronal mass ejections, coronal holes, etc., can significantly alter the interplanetary environment and, consequently, the propagation of cosmic rays in the heliosphere. Understanding the solar modulation process is essential for predicting space weather events that can impact technological systems and human activities in space.

The Parker transport equation, developed by Dr. Eugene Parker, is a cornerstone in the theoretical framework for describing the propagation of GCRs and simulating their fluxes near Earth~\cite{parker1965passage}. The equation is expressed as follows,
\begin{equation}
\!\!\!\!\frac{\partial f}{\partial t}\! =\! -(\bm{V}_{\mathrm{sw}}\!+\!\bm{V}_\mathrm{d}) \cdot\! \nabla f  +\nabla \! \cdot \!\left(\! \stackrel{\leftrightarrow}{K_s} \!\cdot \!\nabla f \!\right) + \frac{1}{3} (\nabla \cdot \bm{V}_\mathrm{sw}) \frac{\partial f}{\partial \ln p},
\end{equation}
where $f(\bm{r}, p, t)$ is the GCR phase-space distribution function, $\bm{r}$ is the heliocentric position, $p$ is the particle momentum, and $t$ is time. In this equation, $\bm{V}_{\mathrm{sw}}$ is the solar wind velocity, $\bm{V}_\mathrm{d}$ is the pitch-angle-averaged drift velocity, and $\stackrel{\leftrightarrow}{K_s}$ represents the diffusion tensor. The Parker transport equation accounts for the essential physical processes that affect cosmic-ray transport, including convection caused by the outward-moving solar wind, particle drift due to large-scale gradients, curvatures, and changes in the heliospheric magnetic field, spatial diffusion resulting from particle interactions with irregularities in the magnetic field, and adiabatic energy losses experienced by cosmic rays as they interact with the expanding solar wind. It provides a comprehensive picture of how cosmic rays are scattered and transported through the heliosphere, ultimately reaching Earth. The Parker transport model incorporates a range of solar activity parameters, including the near-Earth solar magnetic field, solar wind velocity, solar polarity, and the tilt angle of the heliospheric current sheet (HCS), among others. However, traditional transport models based on the Parker transport equation face several significant limitations. These include computational challenges in solving the full equation for multiple particle species, limited adaptability to rapidly changing solar parameters, and difficulties in accurately modeling the time-dependent and highly variable three-dimensional global heliospheric environment, particularly the diffusion and drift coefficients, which are notoriously challenging to determine precisely~\cite{song2021numerical}. Such limitations significantly hinder the ability of traditional transport models to provide accurate predictions of cosmic ray flux.


Recent advances in machine learning, particularly deep learning, offer powerful tools for addressing complex, data-intensive problems in physics and astrophysics~\cite{carrasquilla2017machine,ch2017machine,pang2018equation,carleo2019machine,du2020identifying,Du:2020pmp,du2021jet,zhou2019regressive,shi2022heavy,boehnlein2022colloquium,vanderplas2012introduction,george2018deep}. In both solar activity~\cite{qahwaji2007automatic,colak2009automated,voyant2017machine,malinovic2023application,kasapis2024predicting} and cosmic ray~\cite{paschalis2013artificial,erdmann2018deep,lin2019investigating,schichtel2019constraining,zhang2020deepcr,acciarri2021cosmic,mandrikova2022hybrid,bister2022inference,tsai2022inverting,alvarado2023cosmic,hachaj2023fast,kuznetsov2024energy} research, machine learning has led to significant advancements in modeling and forecasting. These approaches offer new avenues for understanding the relationship between solar phenomena and cosmic-ray flux modulation~\cite{belen2024exploring,polatoglu2024observation}. Deep learning methods also offer alternative solutions to challenges encountered by traditional transport models, enhancing predictive capability where conventional models may have limitations.

\begin{figure*}[tbh!]
\centering
\includegraphics[width=0.75\textwidth]{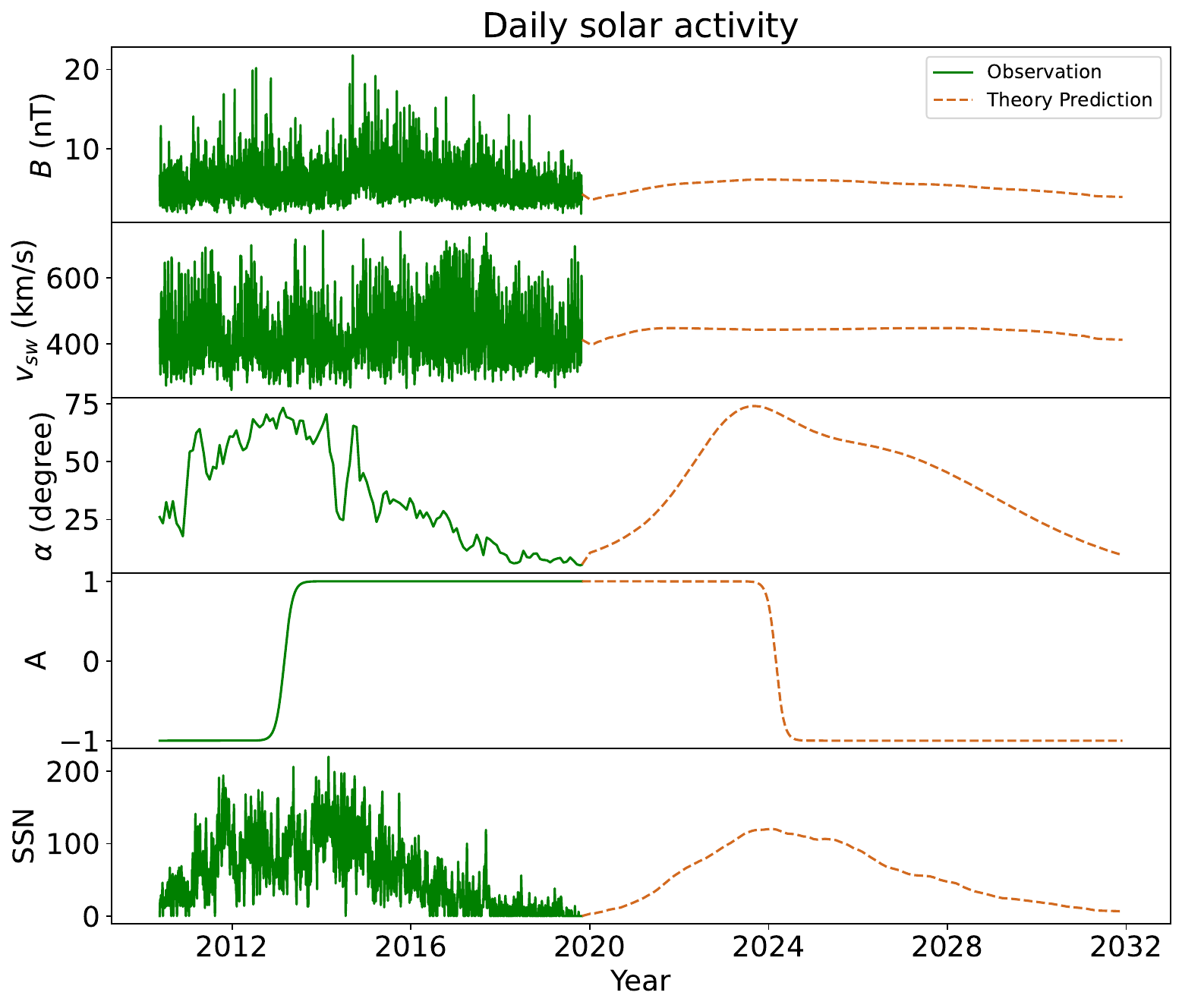}
\caption{Daily solar activity observed from 2010 to 2019 and predicted from 2019 to 2031 are shown as inputs to the LSTM neural network, plotted over time. The parameters displayed include the heliospheric magnetic field (HMF) ($B$), solar wind speed ($v_{sw}$) near Earth, heliospheric current sheet (HCS) tilt angle ($\alpha$), solar polarity (A), and sunspot numbers (SSN).}
\label{fig: Daily_solar_activities_overall} 
\end{figure*}

In this Letter, we aim to address the challenges of forecasting the galactic cosmic-ray spectra near Earth by employing deep learning techniques to establish a direct link between solar activity and the GCRs spectra as measured by Alpha Magnetic Spectrometer (AMS). Focusing on hydrogen and helium, which constitute over 99\% of GCRs~\cite{simpson1983elemental,gaisser1982cosmic}, our model leverages neural networks to discover long-term, hidden correlations within historical data, thereby enhancing the predictive capabilities of our models. This approach bypasses the need for a detailed understanding of the complex transport processes, focusing instead on the empirical relationships between solar activity and cosmic ray flux. Our methodology involves training a neural network on a dataset that integrates a broad range of solar activity parameters and historical cosmic ray flux measurements. This dataset, while limited to the AMS data range, is curated to cover a broad spectrum of solar activity variations and their potential impacts on cosmic ray flux. Key inputs, such as sunspot number (SSN)—a known indicator of solar activity correlated with cosmic-ray modulations~\cite{forbush1954world,forbush1958cosmic,cliver2001coronal,mishra2006cosmic,ross2019behaviour,koldobskiy2022time}—and additional cosmic ray flux data, are incorporated to help the model better capture the nuances of solar--cosmic-ray interactions.

This data-driven approach holds promise for improving cosmic-ray flux prediction accuracy, with potential applications in aerospace, satellite operations, and telecommunications, where reliable forecasts can mitigate the effects of space weather.

\section{Long Short-term Memory Neural network and data preparation}
\label{sec:LSTM}
In this study, we utilize Long Short-Term Memory (LSTM) networks~\cite{hochreiter1997long}, a type of recurrent neural network (RNN) well-suited for handling sequential data~\cite{hua2019deep,tealab2018time,sak2014long}. LSTM networks effectively address the vanishing gradient problem~\cite{basodi2020gradient}, which often impedes traditional RNNs in learning long-term dependencies. Through memory cells and a gating mechanism—including input, output, and forget gates—the LSTM architecture selectively retains or discards information over time, enabling it to capture the complex temporal dynamics of cosmic-ray flux, which are modulated by evolving solar activity patterns.

We adopt a sliding window technique to capture temporal trends in solar activity data, enabling both current and future cosmic-ray flux predictions. Specifically, we apply a 365-day sliding window, shifting by one day, to establish a continuous temporal context for each prediction step. The one-year input data period is selected for its relevance to solar modulation of cosmic rays, capturing both short-term variations and broader trends. For single rigidity-bin flux predictions (where rigidity refers to the momentum-to-charge ratio of cosmic rays, determining their deflection by magnetic fields), we configure the LSTM model with 64 neurons in the hidden layer, while for full-spectrum or multi-species predictions, we use 128 neurons. To prevent overfitting, we incorporate dropout and recurrent dropout rates of 0.05 or 0.40 depending on whether the historical data of cosmic-rays fluxes are included as input, where dropout randomly deactivates a portion of neurons during training to encourage more generalized pattern learning~\cite{srivastava2014dropout,semeniuta2016recurrent}. Each hidden layer neuron functions as a memory cell, processing incoming data, retaining pertinent information, and propagating insights to subsequent time steps, thereby enhancing the network's ability to capture intricate temporal patterns.

The model is trained with a learning rate of 0.0001 for over 1000 epochs to ensure steady convergence to optimal parameter values. We employ the Adamax optimizer~\cite{kingma2014adam}, which is well-suited for handling sparse gradients encountered in LSTM networks. To evaluate performance, we apply the Huber loss function~\cite{10.1214/aoms/1177703732}, a robust metric that combines the strengths of mean squared error and mean absolute error, maintaining sensitivity across various prediction discrepancies.

\begin{figure*}[tbh!]
\centering
\includegraphics[width=0.95\textwidth]{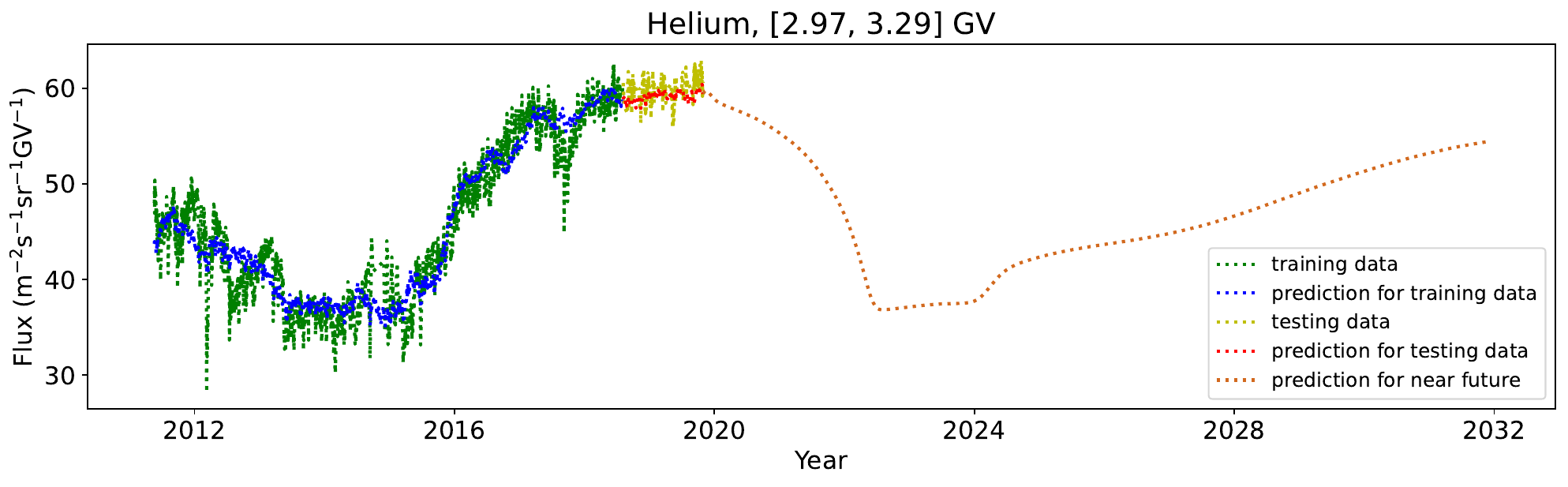}
\caption{Training, testing, and prediction results of Helium flux at [2.97, 3.29] GV one day ahead with four solar parameters of the past year as input. The green (yellow) dashed curve depicts the training (testing) data from AMS measurements while the blue (red) dashed curve depicts the prediction of the LSTM neural network for the training (testing) data. The chocolate dashed curve depicts the prediction of the LSTM neural network for the near future.}
\label{fig: Flux_of_Helium_4_solar} 
\end{figure*}

Our model ingests a comprehensive set of solar activity data spanning the past year, including measurements of the heliospheric magnetic field (HMF) and solar wind speed near Earth~\footnote{\href{https://spdf.gsfc.nasa.gov/pub/data/omni/low_res_omni/}{spdf.gsfc.nasa.gov}
}, HCS tilt angle~\footnote{\href{http://wso.stanford.edu/Tilts.html}{wso.stanford.edu/Tilts}}, and solar polarity (A)~\footnote{\href{http://wso.stanford.edu/Polar.html}{wso.stanford.edu/Polar}}, along with sunspot numbers (SSN)~\footnote{\href{https://www.sidc.be/SILSO/datafiles}{sidc.be}}. In Fig.~\ref{fig: Daily_solar_activities_overall}, we present daily measurements of these solar activity parameters from May 20, 2010, to October 29, 2019, along with predictions extending to December 1, 2031 from the Similar Cycle method~\cite{PredictionOnsetSolarCycle24,ssncycle24,solarcycle25,miao2020prediction}. To achieve daily data resolution, we apply linear interpolation to the raw data, including both the observed HCS tilt angle and predicted solar activity. Additionally, we smooth the solar polarity (A) using a Sigmoid function applied in the reverse interval to ensure smooth predictions. For the AMS cosmic-ray flux data covering 3,085 days from May 20, 2011, to October 29, 2019~\cite{aguilar2022properties}, we allocated 85\% for training (2,635 days) and 15\% for testing (450 days). Prior to training, each solar parameter and cosmic-ray flux value is normalized to the range [0,1] using the min-max scaling, and we set a batch size of 64 to optimize the learning process.

\section{Cosmic-Ray Spectra Prediction with Comparative Setups}

In this study, we investigate five distinct input-output configurations to train our LSTM neural network model, each aimed at evaluating the effects of solar parameters and cosmic-ray fluxes on prediction accuracy. These setups vary in terms of the historical data used, the inclusion of specific solar and cosmic ray indicators, and the prediction goals, e.g., flux of specific elements or the entire spectrum. Below, we outline each configuration and its corresponding results.

We first used inputs mirroring those in the Parker transport model, incorporating observed heliospheric magnetic field (HMF) and solar wind speed near Earth, heliospheric current sheet (HCS) tilt angle, and solar polarity (A) over a one-year period. Our primary goal is to predict the Helium flux in the low rigidity range of [2.97, 3.29] GV one day in advance. With this configuration, the results of the training, testing, and prediction are displayed in Fig.~\ref{fig: Flux_of_Helium_4_solar}. The green dashed curve represents the training data, while the yellow dashed curve shows the testing data. The blue dashed curve corresponds to the LSTM model's predictions for the training data, and the red dashed curve shows the model's predictions for the testing data. By the end of the training process, the LSTM network achieved a mean relative error of 4.65\% on the training data and 1.69\% on the testing data, attributed to the lower variability in the testing set. We confirmed the model’s robustness, as no overfitting was detected. This setup provides a solid baseline for future analyses, though it does not capture some finer features in the training data. We also extended the prediction into the near future, covering solar cycle 25, as shown by the chocolate dashed curve.

\begin{table}[h]
    \centering
    \begin{tabular}{|>{\centering\arraybackslash}p{2.6cm}|>{\centering\arraybackslash}p{2.6cm}|>{\centering\arraybackslash}p{2.6cm}|}
        \hline
       Shuffled Solar Parameters  & Training Data Error & Testing Data Error\\
        \hline
         None  &  4.62\% & 1.61\% \\
        \hline
         $B$ & 4.85\% & 1.81\% \\
        \hline
        $v_{sw}$  & 5.17\% & 1.71\% \\
        \hline
         SSN  & 5.44\% & 8.46\% \\
        \hline
        A  & 7.05\% & 2.94\% \\
        \hline
        $\alpha$  & 8.97\% & 10.78\% \\
        \hline
        All   & 22.4\% & 23.2\% \\
        \hline
    \end{tabular}
    \caption{Mean relative error with different solar parameters shuffled as input to LSTM, including cases where none, one, or all inputs are randomized.}
    \label{tab:permutation}
\end{table}

\begin{figure*}[tbh!]
\centering
\includegraphics[width=0.9\textwidth]{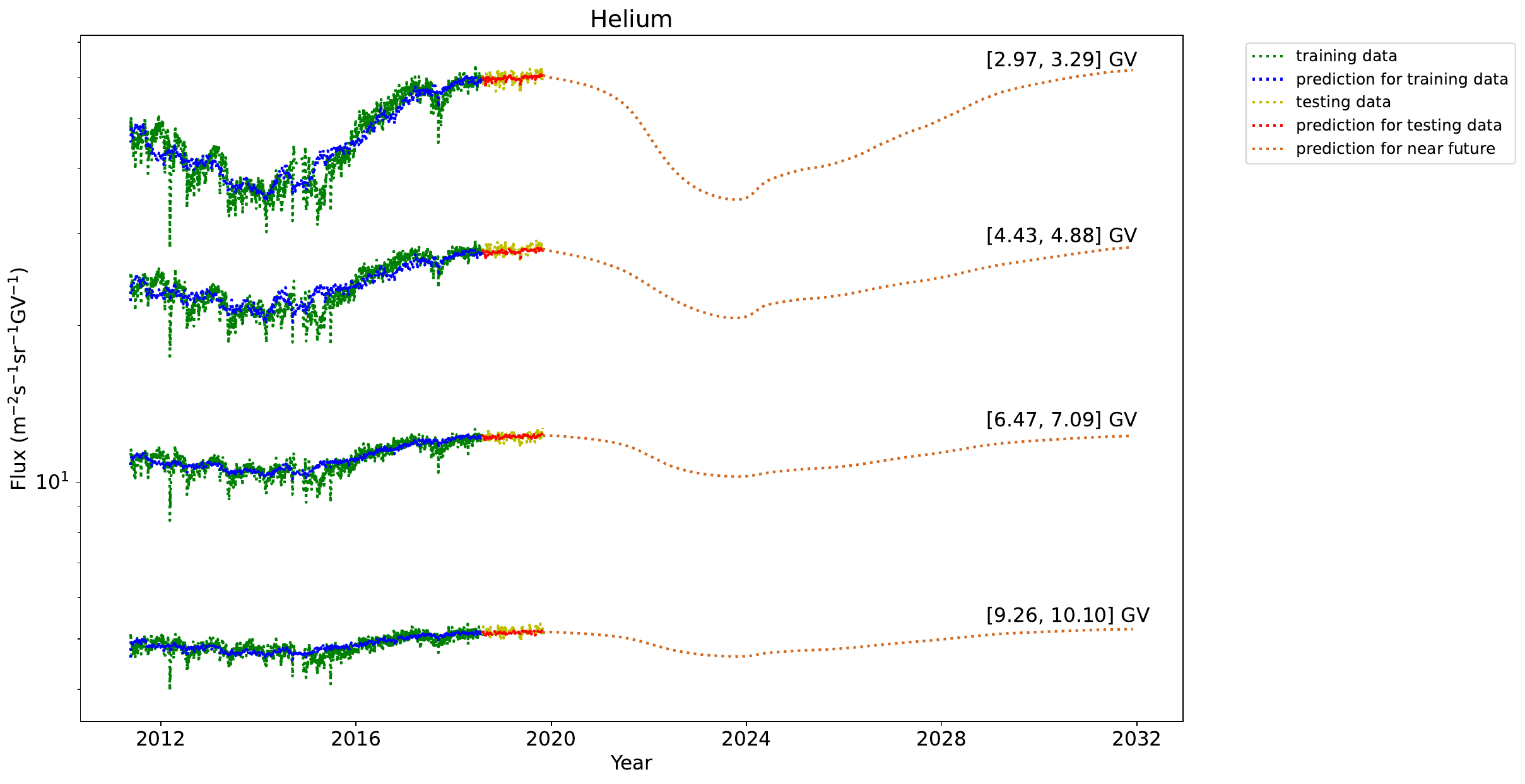}
\caption{Training, testing, and prediction results of Helium spectra one day ahead with five solar parameters and Helium spectra of the past year as input. The green (yellow) dashed curve depicts the training (testing) data from AMS measurements while the blue (red) dashed curve depicts the prediction of the LSTM neural network for the training (testing) data. The chocolate dashed curve depicts the prediction of the LSTM neural network for the near future.}
\label{fig: Spectra_of_Helium_5_solar_He} 
\end{figure*}

The flexibility of neural networks allows us to incorporate additional solar activity parameters, such as sunspot numbers (SSN), as input to further enhance prediction performance. Using the same training approach, this addition reduced mean relative error to 4.62\% for training data and 1.61\% for testing data, illustrating a slight accuracy improvement from incorporating SSN. We further extended the prediction into the near future; the results are provided in the Appendix. The predicted trend aligns well with the results shown in Fig.~\ref{fig: Flux_of_Helium_4_solar}, confirming the model's robustness and the consistency of its predictions with solar activity patterns.
To quantify the contribution of each solar activity parameter, we used permutation feature importance by individually shuffling each parameter's values at random. The resulting error increases reveal the unique contribution of each parameter, shown in Tab.~\ref{tab:permutation}. Based on the diverse training data, the solar parameters' importance, ranked in ascending order, is as follows: HMF ($B$), solar wind speed ($v_{sw}$), SSN, solar polarity (A), and HCS tilt angle ($\alpha$). 

To further improve the prediction performance, we incorporated the past one-year flux of Helium as an additional input, a common and effective practice in time series forecasting~\cite{lam2023learning}. For the initial year, when Helium flux data was unavailable, we assigned values of zero to maintain consistent data size. To mitigate error accumulation during training, we increased both the dropout and recurrent dropout rates to 0.40 in the LSTM neural network, effectively discarding 40\% of the input and neuron connections at each training iteration.
We first stick to predicting the Helium flux within the [2.97, 3.29] GV rigidity range, with training, testing, and prediction results displayed in the Appendix. This configuration allowed the model to capture fine structural details in the training data more accurately. The model achieved a mean relative error of 3.34\% on the training data and 1.33\% on the testing data. These results represent a notable improvement in prediction accuracy over previous models, highlighting the effectiveness of incorporating historical flux data and increasing dropout rates. Notably, in the latter phase of the near-future prediction, the model showcases a more pronounced upward trend.

\begin{figure*}[tbh!]
\centering
\includegraphics[width=0.75\textwidth]{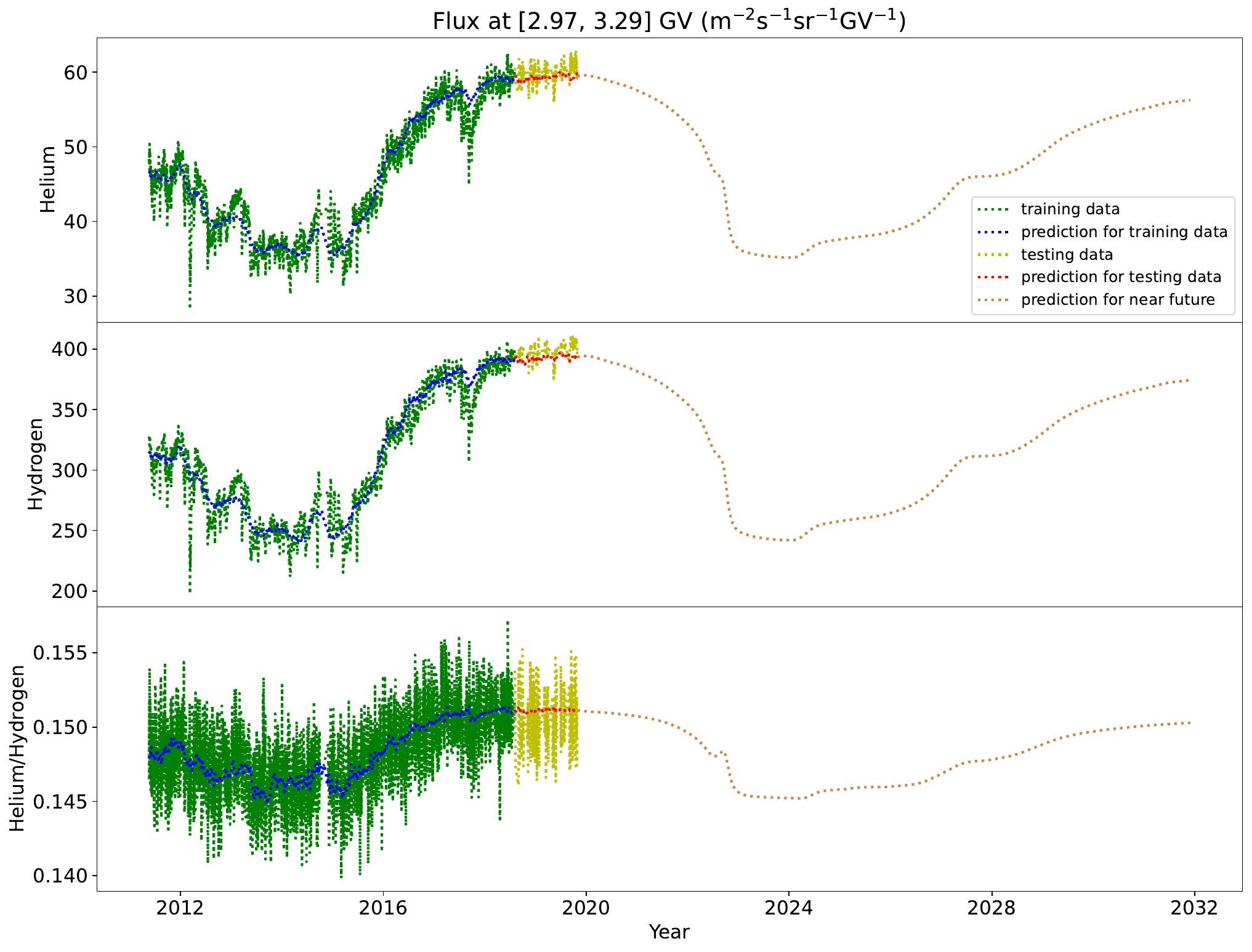}
\caption{Training, testing, and prediction results of Helium and Hydrogen flux as well as their ratio at [2.97, 3.29] GV one day ahead with five solar parameters of the past year as input. The green (yellow) dashed curve depicts the training (testing) data from AMS measurements while the blue (red) dashed curve depicts the prediction of the LSTM neural network for the training (testing) data. The chocolate dashed curve depicts the prediction of the LSTM neural network for the near future.}
\label{fig: Flux_of_Helium_Hydrogen_5_solar} 
\end{figure*}

\begin{figure*}[tbh!]
\centering
\includegraphics[width=0.95\textwidth]{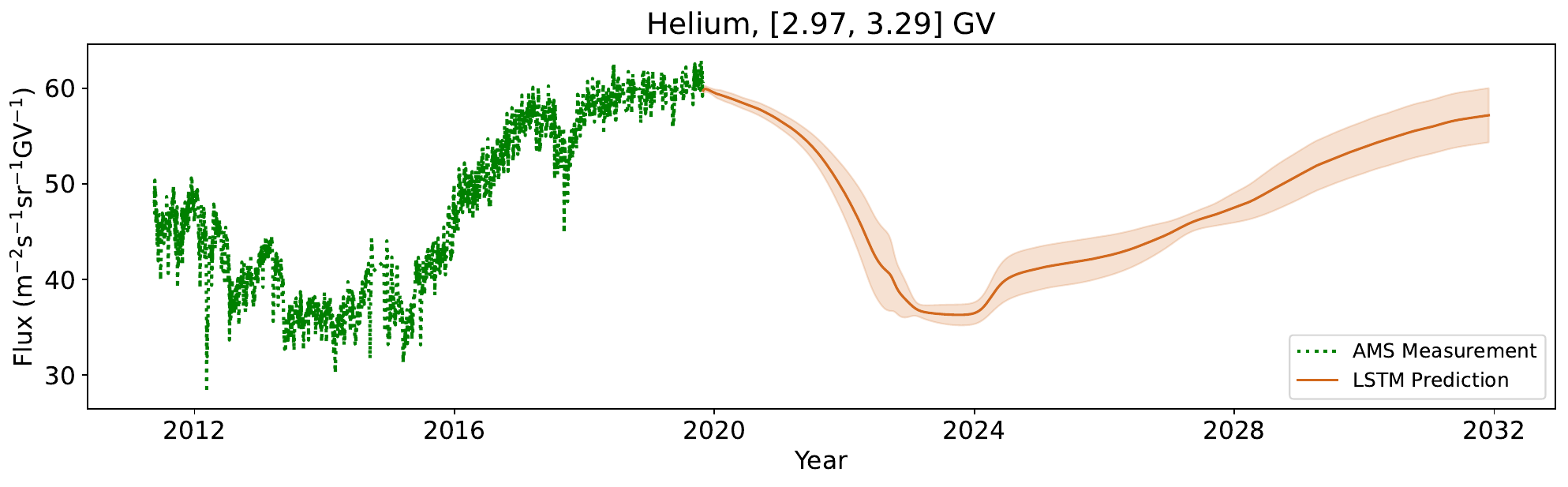}
\caption{The green dashed curve depicts the AMS measurements for Helium flux at [2.97, 3.29] GV, while the chocolate-colored band shows the LSTM neural network's near-future prediction, averaged across different model configurations.}
\label{fig: Flux_of_Helium_Fill_Between} 
\end{figure*}

Moving from a single-bin Helium flux prediction, we also extended this framework to predict the full Helium spectrum, using a similar LSTM network with an increased neuron count (128) to accommodate the task’s complexity. Incorporating the past one-year Helium spectra as the additional input, results for four selected rigidity bins are presented in Fig.~\ref{fig: Spectra_of_Helium_5_solar_He}, indicating that the model successfully predicts the entire spectrum. For the [2.97, 3.29] GV rigidity bin, the model achieved a mean relative error of 5.27\% for the training data and 1.29\% for the testing data. This accuracy for the specific bin is lower than those of the earlier single-bin predictions, reflecting the trade-off in precision when expanding the model to encompass the entire spectral range. Additionally, we note that the mean relative error tends to decrease as rigidity increases.

Furthermore, this framework is highly generalizable for predicting the fluxes or the full spectra of multiple cosmic-ray elements simultaneously. For instance, we retrained the LSTM neural network from scratch using five solar parameters of the past year to concurrently predict the fluxes of both Hydrogen and Helium in the [2.97, 3.29] GV range. The model achieved a mean relative error of 3.79\% for the training data and 1.64\% for the testing data for Helium, and 3.39\% and 1.64\% for Hydrogen, respectively. The training, testing, and prediction results for the fluxes of Helium, Hydrogen, and their ratio are presented in Fig.~\ref{fig: Flux_of_Helium_Hydrogen_5_solar}, further demonstrating the model's robustness and cost-effectiveness across different elements. Notably, the helium-to-hydrogen flux ratio trends align with those of the hydrogen flux.

Lastly, to address discrepancies among model setups in Helium flux prediction in the [2.97, 3.29] GV rigidity range, we averaged the results from each setup and quantified uncertainty using their root mean square deviation. The averaged prediction, along with the associated uncertainty, is provided in Fig.~\ref{fig: Flux_of_Helium_Fill_Between}.

\section{Conclusions} 
\label{sec: Conclusions}
In this work, we have established a novel and effective correspondence between historical solar activity and near-Earth cosmic ray flux using Long Short-Term Memory (LSTM) networks, a powerful deep learning technique. By applying permutation feature importance, we evaluated each solar parameter’s specific impact on model accuracy, offering insights into the distinct roles of each factor in cosmic ray modulation. This framework facilitates the integration of diverse solar activity data, such as sunspot numbers, and lays the groundwork for incorporating an even broader range of observational inputs to improve predictive performance. Compared to traditional transport models, which are limited in integrating varied datasets, our deep learning approach demonstrates superior flexibility and adaptability.

Including historical cosmic ray flux data as additional input significantly improves the accuracy of short-term, one-day-ahead predictions. This capability is valuable for space weather forecasting, where precise and timely predictions are essential to mitigate cosmic ray effects on human health and technology. For long-term predictions, we manage error accumulation by applying a higher dropout rate during LSTM training, preserving model stability in extended forecasting. By exploring various modeling schemes, our framework demonstrates the ability to reliably predict cosmic ray spectra across different species and forecast durations, from daily to multi-year forecasts. 

In summary, these findings highlight the capacity of machine learning to enhance or even surpass traditional physics-based models in certain aspects, offering a scalable, data-driven solution with high forecasting utility. This approach not only improves the precision of cosmic ray forecasting but also lays a foundation for future improvements, including the potential for integrating additional solar and cosmic ray data as they become available. Overall, this study demonstrates the feasibility and promise of deep learning for cosmic ray flux prediction, offering an advanced pathway to support space weather monitoring, technological safeguarding, and future space mission planning.

\section*{Acknowledgments}
We appreciate the insightful discussions with Weiwei Xu, Ran Huo, Kai Zhou, Zhaoming Wang, Yao Chen, and Siqi Wang. This work is supported by the Taishan Scholars Program of Shandong Province under Grant No. tsqnz20221162 (Y. D.) and Grant No. 202103143 (X. L.), the Shandong Excellent Young Scientists Fund Program (Overseas) under Grant No. 2023HWYQ-106 (Y. D.), the NSFC under Grant No. U2106201 (X. L.) and Grant No. 42404172 (X. S.), and the Natural Science Foundation of Shandong Province under Grant No. ZR2022QD033 (X. S.). 

\section*{appendix}
\begin{figure*}[tbh!]
\centering
\includegraphics[width=0.95\textwidth]{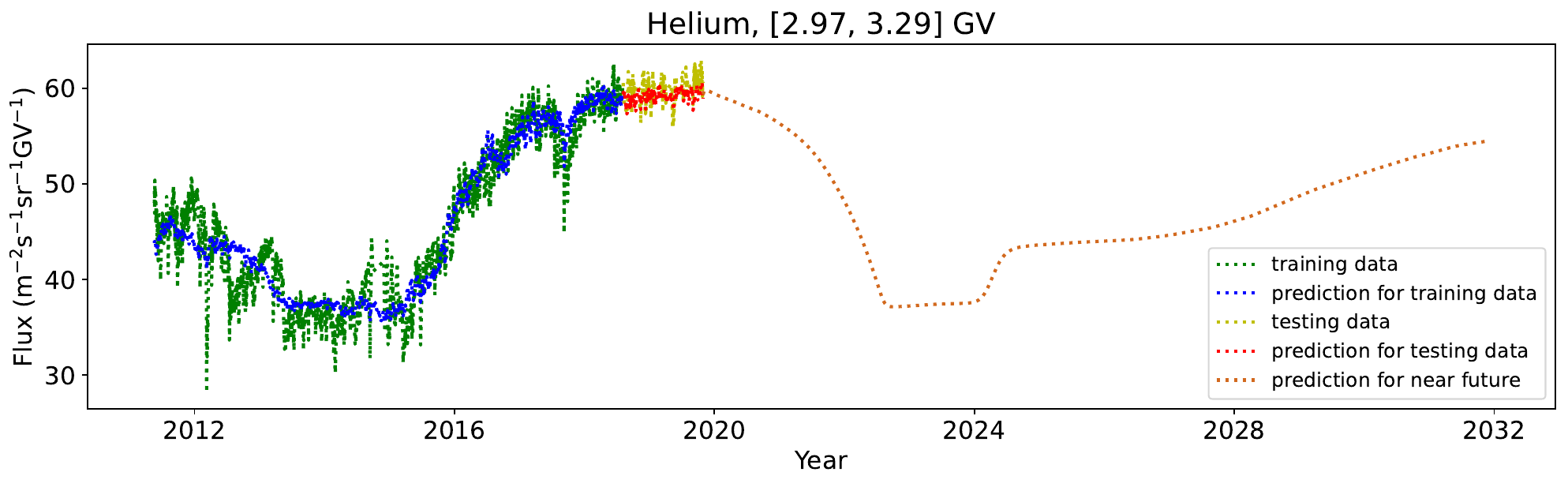}
\caption{Training, testing, and prediction results of Helium flux at [2.97, 3.29] GV one day ahead with five solar parameters of the past year as input. The green (yellow) dashed curve depicts the training (testing) data from AMS measurements while the blue (red) dashed curve depicts the prediction of the LSTM neural network for the training (testing) data. The chocolate dashed curve depicts the prediction of the LSTM neural network for the near future.}
\label{fig: Flux_of_Helium_5_solar} 
\end{figure*}

\begin{figure*}[tbh!]
\centering
\includegraphics[width=0.95\textwidth]{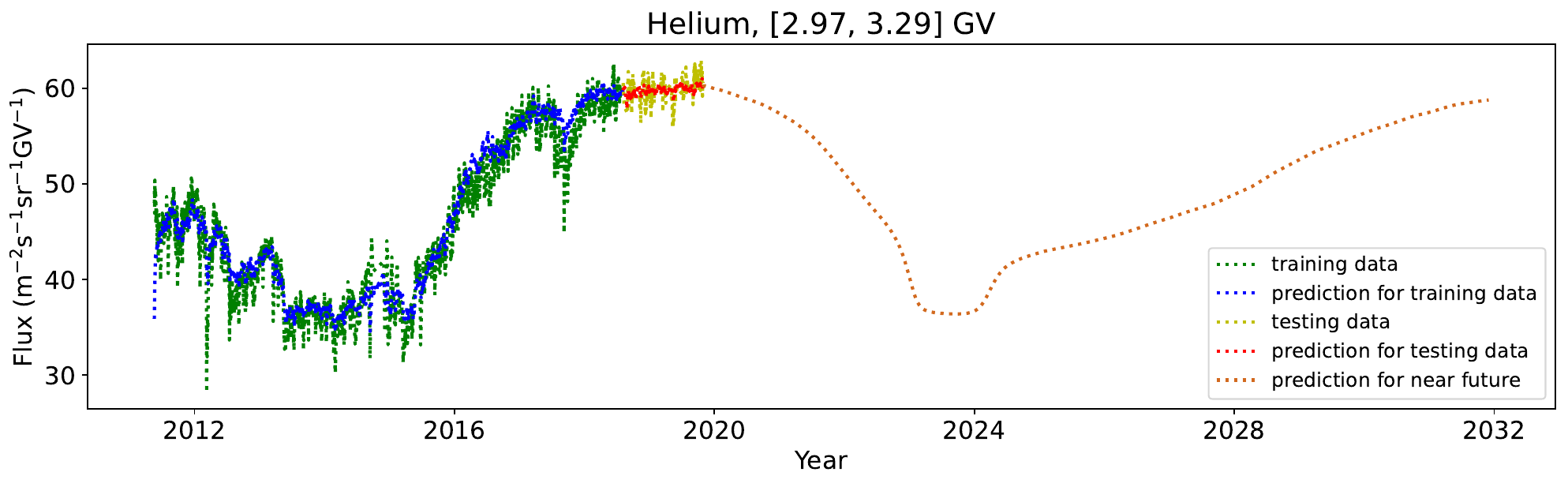}
\caption{Training, testing, and prediction results of Helium flux at [2.97, 3.29] GV one day ahead with five solar parameters and Helium flux at [2.97, 3.29] GV of the past year as input. The green (yellow) dashed curve depicts the training (testing) data from AMS measurements while the blue (red) dashed curve depicts the prediction of the LSTM neural network for the training (testing) data. The chocolate dashed curve depicts the prediction of the LSTM neural network for the near future.}
\label{fig: Flux_of_Helium_5_solar_He} 
\end{figure*}

Figures.~\ref{fig: Flux_of_Helium_5_solar} and \ref{fig: Flux_of_Helium_5_solar_He} display the training, testing, and prediction results for Helium flux in the [2.97, 3.29] GV rigidity range one day ahead. These figures compare models trained with five solar parameters alone and those augmented with the historical Helium flux as an additional input.

\clearpage


\begin{thebibliography}{}
\expandafter\ifx\csname natexlab\endcsname\relax\def\natexlab#1{#1}\fi
\providecommand{\url}[1]{\href{#1}{#1}}
\providecommand{\dodoi}[1]{doi:~\href{http://doi.org/#1}{\nolinkurl{#1}}}
\providecommand{\doeprint}[1]{\href{http://ascl.net/#1}{\nolinkurl{http://ascl.net/#1}}}
\providecommand{\doarXiv}[1]{\href{https://arxiv.org/abs/#1}{\nolinkurl{https://arxiv.org/abs/#1}}}

\end{thebibliography}


\begin{thebibliography}{99}
\expandafter\ifx\csname natexlab\endcsname\relax\def\natexlab#1{#1}\fi
\providecommand{\url}[1]{\href{#1}{#1}}
\providecommand{\dodoi}[1]{doi:~\href{http://doi.org/#1}{\nolinkurl{#1}}}
\providecommand{\doeprint}[1]{\href{http://ascl.net/#1}{\nolinkurl{http://ascl.net/#1}}}
\providecommand{\doarXiv}[1]{\href{https://arxiv.org/abs/#1}{\nolinkurl{https://arxiv.org/abs/#1}}}

\bibitem[{Acciarri {et~al.}(2021)Acciarri, Adams, Andreopoulos, Asaadi, Babicz,
  Backhouse, Badgett, Bagby, Barker, Basque, {et~al.}}]{acciarri2021cosmic}
Acciarri, R., Adams, C., Andreopoulos, C., {et~al.} 2021, Front. Artif.
  Intell., 4, 649917

\bibitem[{Aguilar {et~al.}(2021)Aguilar, Cavasonza, Ambrosi, Arruda, Attig,
  Barao, Barrin, Bartoloni, Ba{\c{s}}e{\u{g}}mez-du Pree, Battiston,
  {et~al.}}]{aguilar2021periodicities}
Aguilar, M., Cavasonza, L.~A., Ambrosi, G., {et~al.} 2021, Phys. Rev. Lett.,
  127, 271102

\bibitem[{Aguilar {et~al.}(2022)Aguilar, Cavasonza, Ambrosi, Arruda, Attig,
  Barao, Barrin, Bartoloni, Ba{\c{s}}e{\u{g}}mez-du Pree, Battiston,
  {et~al.}}]{aguilar2022properties}
---. 2022, Phys. Rev. Lett., 128, 231102

\bibitem[{Alvarado {et~al.}(2023)Alvarado, Capistr{\'a}n, Torres, Sacahu{\'I},
  \& Alfaro}]{alvarado2023cosmic}
Alvarado, A., Capistr{\'a}n, T., Torres, I., Sacahu{\'I}, J., \& Alfaro, R.
  2023, arXiv:2310.06938

\bibitem[{Basodi {et~al.}(2020)Basodi, Ji, Zhang, \& Pan}]{basodi2020gradient}
Basodi, S., Ji, C., Zhang, H., \& Pan, Y. 2020, Big Data Min. Anal., 3, 196

\bibitem[{Belen {et~al.}(2024)Belen, Lelo{\u{g}}lu, \&
  Demirk{\"o}z}]{belen2024exploring}
Belen, B., Lelo{\u{g}}lu, U., \& Demirk{\"o}z, M. 2024, Adv. Space Res.

\bibitem[{Bister {et~al.}(2022)Bister, Erdmann, K{\"o}the, \&
  Schulte}]{bister2022inference}
Bister, T., Erdmann, M., K{\"o}the, U., \& Schulte, J. 2022, Eur. Phys. J. C,
  82, 1

\bibitem[{Boehnlein {et~al.}(2022)Boehnlein, Diefenthaler, Sato, Schram,
  Ziegler, Fanelli, Hjorth-Jensen, Horn, Kuchera, Lee,
  {et~al.}}]{boehnlein2022colloquium}
Boehnlein, A., Diefenthaler, M., Sato, N., {et~al.} 2022, Rev. Mod. Phys., 94,
  031003

\bibitem[{Carleo {et~al.}(2019)Carleo, Cirac, Cranmer, Daudet, Schuld, Tishby,
  Vogt-Maranto, \& Zdeborov{\'a}}]{carleo2019machine}
Carleo, G., Cirac, I., Cranmer, K., {et~al.} 2019, Rev. Mod. Phys., 91, 045002

\bibitem[{Carrasquilla \& Melko(2017)}]{carrasquilla2017machine}
Carrasquilla, J., \& Melko, R.~G. 2017, Nat. Phys.

\bibitem[{Chen {et~al.}(2023)Chen, Xu, Song, Huo, \& Luo}]{chen2023astronaut}
Chen, X., Xu, S., Song, X., Huo, R., \& Luo, X. 2023, Space Weather, 21,
  e2022SW003285

\bibitem[{Ch'ng {et~al.}(2017)Ch'ng, Carrasquilla, Melko, \&
  Khatami}]{ch2017machine}
Ch'ng, K., Carrasquilla, J., Melko, R.~G., \& Khatami, E. 2017, Phys. Rev. X,
  7, 031038

\bibitem[{Cliver \& Ling(2001)}]{cliver2001coronal}
Cliver, E., \& Ling, A. 2001, ApJ, 556, 432

\bibitem[{Colak \& Qahwaji(2009)}]{colak2009automated}
Colak, T., \& Qahwaji, R. 2009, Space Weather, 7

\bibitem[{Cucinotta {et~al.}(2014)Cucinotta, Alp, Sulzman, \&
  Wang}]{cucinotta2014space}
Cucinotta, F.~A., Alp, M., Sulzman, F.~M., \& Wang, M. 2014, Life Sci. Space
  Res., 2, 54

\bibitem[{de~Wet {et~al.}(2020)de~Wet, Slaba, Rahmanifard, Wilson, Jordan,
  Townsend, Schwadron, \& Spence}]{de2020crater}
de~Wet, W.~C., Slaba, T.~C., Rahmanifard, F., {et~al.} 2020, Life Sci. Space
  Res., 26, 149

\bibitem[{Du {et~al.}(2020{\natexlab{a}})Du, Pablos, \& Tywoniuk}]{Du:2020pmp}
Du, Y.-L., Pablos, D., \& Tywoniuk, K. 2020{\natexlab{a}}, JHEP, 21, 206

\bibitem[{Du {et~al.}(2022)Du, Pablos, \& Tywoniuk}]{du2021jet}
---. 2022, Phys. Rev. Lett., 128, 012301

\bibitem[{Du {et~al.}(2020{\natexlab{b}})Du, Zhou, Steinheimer, Pang,
  Motornenko, Zong, Wang, \& St{\"o}cker}]{du2020identifying}
Du, Y.-L., Zhou, K., Steinheimer, J., {et~al.} 2020{\natexlab{b}}, Eur. Phys.
  J. C, 80, 1

\bibitem[{Erdmann {et~al.}(2018)Erdmann, Glombitza, \& Walz}]{erdmann2018deep}
Erdmann, M., Glombitza, J., \& Walz, D. 2018, Astropart. Phys., 97, 46

\bibitem[{Forbush(1954)}]{forbush1954world}
Forbush, S.~E. 1954, J. Geophys. Res., 59, 525

\bibitem[{Forbush(1958)}]{forbush1958cosmic}
---. 1958, J. Geophys. Res., 63, 651

\bibitem[{Gaisser(1982)}]{gaisser1982cosmic}
Gaisser, T.~K. 1982, Comments Nucl. Part. Phys., 11, 25

\bibitem[{George \& Huerta(2018)}]{george2018deep}
George, D., \& Huerta, E. 2018, Phys. Rev. D, 97, 044039

\bibitem[{Hachaj {et~al.}(2023)Hachaj, Bibrzycki, \&
  Piekarczyk}]{hachaj2023fast}
Hachaj, T., Bibrzycki, {\L}., \& Piekarczyk, M. 2023, IEEE Access, 11, 7410

\bibitem[{Hochreiter(1997)}]{hochreiter1997long}
Hochreiter, S. 1997, Neural Comput.

\bibitem[{Hua {et~al.}(2019)Hua, Zhao, Li, Chen, Liu, \& Zhang}]{hua2019deep}
Hua, Y., Zhao, Z., Li, R., {et~al.} 2019, IEEE Commun. Mag., 57, 114

\bibitem[{Huber(1964)}]{10.1214/aoms/1177703732}
Huber, P.~J. 1964, Ann. Math. Stat., 35, 73

\bibitem[{Kasapis {et~al.}(2024)Kasapis, Kitiashvili, Kosovichev, Stefan, \&
  Apte}]{kasapis2024predicting}
Kasapis, S., Kitiashvili, I.~N., Kosovichev, A.~G., Stefan, J.~T., \& Apte, B.
  2024, arXiv preprint arXiv:2402.08890

\bibitem[{Kingma \& Ba(2014)}]{kingma2014adam}
Kingma, D.~P., \& Ba, J. 2014, arXiv:1412.6980

\bibitem[{Koldobskiy {et~al.}(2022)Koldobskiy, K{\"a}hk{\"o}nen, Hofer,
  Krivova, Kovaltsov, \& Usoskin}]{koldobskiy2022time}
Koldobskiy, S.~A., K{\"a}hk{\"o}nen, R., Hofer, B., {et~al.} 2022, Sol. Phys.,
  297, 38

\bibitem[{K{\'o}ta(2013)}]{kota2013theory}
K{\'o}ta, J. 2013, Space Sci. Rev., 176, 391

\bibitem[{Kuznetsov {et~al.}(2024)Kuznetsov, Petrov, Plokhikh, \&
  Sotnikov}]{kuznetsov2024energy}
Kuznetsov, M.~Y., Petrov, N., Plokhikh, I., \& Sotnikov, V. 2024, J. Cosmol.
  Astropart. Phys., 2024, 125

\bibitem[{Lam {et~al.}(2023)Lam, Sanchez-Gonzalez, Willson, Wirnsberger,
  Fortunato, Alet, Ravuri, Ewalds, Eaton-Rosen, Hu, {et~al.}}]{lam2023learning}
Lam, R., Sanchez-Gonzalez, A., Willson, M., {et~al.} 2023, Science, 382, 1416

\bibitem[{Lin {et~al.}(2019)Lin, Bi, \& Yin}]{lin2019investigating}
Lin, S.-J., Bi, X.-J., \& Yin, P.-F. 2019, Phys. Rev. D, 100, 103014

\bibitem[{Luo {et~al.}(2017)Luo, Potgieter, Zhang, \& Feng}]{luo2017numerical}
Luo, X., Potgieter, M.~S., Zhang, M., \& Feng, X. 2017, ApJ, 839, 53

\bibitem[{Luo {et~al.}(2020)Luo, Zhang, Feng, Potgieter, Shen, \&
  Bazilevskaya}]{luo2020numerical}
Luo, X., Zhang, M., Feng, X., {et~al.} 2020, ApJ, 899, 90

\bibitem[{Luo {et~al.}(2011)Luo, Zhang, Rassoul, \& Pogorelov}]{luo2011cosmic}
Luo, X., Zhang, M., Rassoul, H.~K., \& Pogorelov, N.~V. 2011, ApJ, 730, 13

\bibitem[{Luo {et~al.}(2013)Luo, Zhang, Rassoul, Pogorelov, \&
  Heerikhuisen}]{luo2013galactic}
Luo, X., Zhang, M., Rassoul, H.~K., Pogorelov, N.~V., \& Heerikhuisen, J. 2013,
  ApJ, 764, 85

\bibitem[{Malinovi{\'c}-Mili{\'c}evi{\'c}
  {et~al.}(2023)Malinovi{\'c}-Mili{\'c}evi{\'c}, Radovanovi{\'c},
  Radenkovi{\'c}, Vyklyuk, Milovanovi{\'c}, Milanovi{\'c}~Pe{\v{s}}i{\'c},
  Milenkovi{\'c}, Popovi{\'c}, Petrovi{\'c}, Sydor,
  {et~al.}}]{malinovic2023application}
Malinovi{\'c}-Mili{\'c}evi{\'c}, S., Radovanovi{\'c}, M.~M., Radenkovi{\'c},
  S.~D., {et~al.} 2023, Mathematics, 11, 795

\bibitem[{Mandrikova \& Mandrikova(2022)}]{mandrikova2022hybrid}
Mandrikova, O., \& Mandrikova, B. 2022, Symmetry, 14, 744

\bibitem[{Miao {et~al.}(2015)Miao, Gong, Li, \& Tingling}]{solarcycle25}
Miao, J., Gong, J., Li, Z., \& Tingling, R. 2015, Sci. China- Phys. Mech.
  Astron., 45, 099601

\bibitem[{MIAO {et~al.}(2007)MIAO, WANG, LIU, \&
  GONG}]{PredictionOnsetSolarCycle24}
MIAO, J., WANG, J., LIU, S., \& GONG, J. 2007, CJSS, 27, 448

\bibitem[{Miao {et~al.}(2020)Miao, Wang, Ren, \& Li}]{miao2020prediction}
Miao, J., Wang, X., Ren, T.-L., \& Li, Z.-T. 2020, Res. Astron. Astrophys., 20,
  004

\bibitem[{Mishra {et~al.}(2006)Mishra, Gupta, \& Mishra}]{mishra2006cosmic}
Mishra, A., Gupta, M., \& Mishra, V. 2006, Sol. Phys., 239, 475

\bibitem[{O'Neill {et~al.}(2015)O'Neill, Golge, \& Slaba}]{o2015badhwar}
O'Neill, P., Golge, S., \& Slaba, T. 2015, Badhwar-O'Neill 2014 galactic cosmic
  ray flux model description, Tech. rep.

\bibitem[{Pang {et~al.}(2018)Pang, Zhou, Su, Petersen, St{\"o}cker, \&
  Wang}]{pang2018equation}
Pang, L.-G., Zhou, K., Su, N., {et~al.} 2018, Nature Commun., 9, 210

\bibitem[{Parker(1965)}]{parker1965passage}
Parker, E.~N. 1965, Planet. Space Sci., 13, 9

\bibitem[{Paschalis {et~al.}(2013)Paschalis, Sarlanis, \&
  Mavromichalaki}]{paschalis2013artificial}
Paschalis, P., Sarlanis, C., \& Mavromichalaki, H. 2013, Sol. Phys., 282, 303

\bibitem[{Polatoglu(2024)}]{polatoglu2024observation}
Polatoglu, A. 2024, IJCESEN, 10

\bibitem[{Potgieter(2014)}]{potgieter2014very}
Potgieter, M. 2014, Braz. J. Phys., 44, 581

\bibitem[{Potgieter(2017)}]{potgieter2017global}
---. 2017, Adv. Space Res., 60, 848

\bibitem[{Potgieter(2013)}]{potgieter2013solar}
Potgieter, M.~S. 2013, Living Rev. Sol. Phys., 10, 3

\bibitem[{Qahwaji \& Colak(2007)}]{qahwaji2007automatic}
Qahwaji, R., \& Colak, T. 2007, Sol. Phys., 241, 195

\bibitem[{Quenby(1984)}]{quenby1984theory}
Quenby, J. 1984, Space Sci. Rev., 37, 201

\bibitem[{Ross \& Chaplin(2019)}]{ross2019behaviour}
Ross, E., \& Chaplin, W.~J. 2019, Sol. Phys., 294, 1

\bibitem[{Sak {et~al.}(2014)Sak, Senior, \& Beaufays}]{sak2014long}
Sak, H., Senior, A., \& Beaufays, F. 2014, in Fifteenth annual conference of
  the international speech communication association

\bibitem[{Schichtel {et~al.}(2019)Schichtel, Spannowsky, \&
  Waite}]{schichtel2019constraining}
Schichtel, P., Spannowsky, M., \& Waite, P. 2019, EPL, 127, 61002

\bibitem[{Semeniuta {et~al.}(2016)Semeniuta, Severyn, \&
  Barth}]{semeniuta2016recurrent}
Semeniuta, S., Severyn, A., \& Barth, E. 2016, arXiv preprint arXiv:1603.05118

\bibitem[{Shen {et~al.}(2019)Shen, Qin, Zuo, \& Wei}]{shen2019modulation}
Shen, Z.-N., Qin, G., Zuo, P., \& Wei, F. 2019, ApJ, 887, 132

\bibitem[{Shi {et~al.}(2022)Shi, Zhou, Zhao, Mukherjee, \&
  Zhuang}]{shi2022heavy}
Shi, S., Zhou, K., Zhao, J., Mukherjee, S., \& Zhuang, P. 2022, Phys. Rev. D,
  105, 014017

\bibitem[{Simonsen {et~al.}(2020)Simonsen, Slaba, Guida, \&
  Rusek}]{simonsen2020nasa}
Simonsen, L.~C., Slaba, T.~C., Guida, P., \& Rusek, A. 2020, PLOS Biol., 18,
  e3000669

\bibitem[{Simpson(1983)}]{simpson1983elemental}
Simpson, J. 1983, Annu. Rev. Nucl. Sci., 33

\bibitem[{Simpson(1998)}]{simpson1998brief}
---. 1998, Space Sci. Rev., 83, 169

\bibitem[{Song {et~al.}(2021)Song, Luo, Potgieter, Liu, \&
  Geng}]{song2021numerical}
Song, X., Luo, X., Potgieter, M.~S., Liu, X., \& Geng, Z. 2021, ApJS, 257, 48

\bibitem[{Srivastava {et~al.}(2014)Srivastava, Hinton, Krizhevsky, Sutskever,
  \& Salakhutdinov}]{srivastava2014dropout}
Srivastava, N., Hinton, G.~E., Krizhevsky, A., Sutskever, I., \& Salakhutdinov,
  R. 2014, J. Mach. Learn. Res., 15, 1929

\bibitem[{Tealab(2018)}]{tealab2018time}
Tealab, A. 2018, FCIJ, 3, 334

\bibitem[{Tsai {et~al.}(2022)Tsai, Chung, Yuan, \& Cheung}]{tsai2022inverting}
Tsai, Y.-L.~S., Chung, Y.-L., Yuan, Q., \& Cheung, K. 2022, J. Cosmol.
  Astropart. Phys., 2022, 044

\bibitem[{VanderPlas {et~al.}(2012)VanderPlas, Connolly, Ivezi{\'c}, \&
  Gray}]{vanderplas2012introduction}
VanderPlas, J., Connolly, A.~J., Ivezi{\'c}, {\v{Z}}., \& Gray, A. 2012, in
  2012 conference on intelligent data understanding, IEEE, 47--54

\bibitem[{Vos \& Potgieter(2016)}]{vos2016global}
Vos, E., \& Potgieter, M. 2016, Sol. Phys., 291, 2181

\bibitem[{Vos \& Potgieter(2015)}]{vos2015new}
Vos, E.~E., \& Potgieter, M.~S. 2015, ApJ, 815, 119

\bibitem[{Voyant {et~al.}(2017)Voyant, Notton, Kalogirou, Nivet, Paoli, Motte,
  \& Fouilloy}]{voyant2017machine}
Voyant, C., Notton, G., Kalogirou, S., {et~al.} 2017, Renew. Energy, 105, 569

\bibitem[{Wang {et~al.}(2008)Wang, Miao, Liu, Gong, \& Cuilian}]{ssncycle24}
Wang, J., Miao, J., Liu, S., Gong, J., \& Cuilian, Z. 2008, Sci. China- Phys.
  Mech. Astron., 38, 1097

\bibitem[{Zhang \& Bloom(2020)}]{zhang2020deepcr}
Zhang, K., \& Bloom, J.~S. 2020, ApJ, 889, 24

\bibitem[{Zhou {et~al.}(2019)Zhou, Endr{\H o}di, Pang, \&
  St{\"o}cker}]{zhou2019regressive}
Zhou, K., Endr{\H o}di, G., Pang, L.-G., \& St{\"o}cker, H. 2019, Phys. Rev.,
  D100, 011501

\end{thebibliography}

\end{document}